\documentclass[11pt]{article}
\usepackage{fa2025}
\usepackage{amsmath}
\usepackage{cite}
\usepackage{url}
\usepackage{graphicx}
\usepackage{color}
\usepackage{siunitx}
\usepackage[utf8]{inputenc}

\usepackage{amsfonts}
\usepackage{amssymb}
\usepackage{algorithmic}
\usepackage{algorithm}
\usepackage{balance}
\usepackage{booktabs}
\usepackage{bm}
\usepackage{cancel}
\usepackage{caption}
\usepackage{dirtytalk}
\usepackage{lipsum}
\usepackage{mathrsfs}
\usepackage{mathtools}
\usepackage{microtype}
\usepackage{multirow}
\usepackage{nicematrix}
\usepackage{nicefrac}
\usepackage{optidef}
\usepackage{pifont}
\usepackage{soul}
\usepackage{stfloats}
\usepackage[labelformat=parens,subrefformat=parens]{subfig}
\usepackage{textcomp}
\usepackage{upgreek}
\usepackage{verbatim}

\usepackage[bookmarks=false, colorlinks=true, allcolors=blue]{hyperref}
\usepackage[nameinlink,capitalise]{cleveref}

\makeatletter
\newcommand{\dotDelta}{{\vphantom{\Delta}\mathpalette\d@tD@lta\relax}}
\newcommand{\d@tD@lta}[2]{%
  \ooalign{\hidewidth$\m@th#1\mkern-1mu\cdot$\hidewidth\cr$\m@th#1\Delta$\cr}%
}
\makeatother

\title{Spatially Selective Active Noise Control for Open-fitting Hearables with Acausal Optimization}

\multauthor
{Tong Xiao$^{*}$ \hspace{1cm} Simon Doclo } 
{ 
Department of Medical Physics and Acoustics and Cluster of Excellence Hearing4all, \\ Carl von Ossietzky Universit\"{a}t Oldenburg, Germany
\correspondingauthor{Tong.Xiao@uni-oldenburg.de}{Tong Xiao and Simon Doclo.}
}

\begin{document}

\maketitle
\begin{abstract}
Recent advances in active noise control have enabled the development of hearables with spatial selectivity, which actively suppress undesired noise while preserving desired sound from specific directions. In this work, we propose an improved approach to spatially selective active noise control that incorporates acausal relative impulse responses into the optimization process, resulting in significantly improved performance over the causal design. We evaluate the system through simulations using a pair of open-fitting hearables with spatially localized speech and noise sources in an anechoic environment. Performance is evaluated in terms of speech distortion, noise reduction, and signal-to-noise ratio improvement across different delays and degrees of acausality. Results show that the proposed acausal optimization consistently outperforms the causal approach across all metrics and scenarios, as acausal filters more effectively characterize the response of the desired source.
\end{abstract}
\keywords{active noise control, spatial selectivity, causality, beamforming}

\section{Introduction}\label{sec:introduction}
\begin{figure}[t]
    \centering
    \includegraphics[width=0.99\linewidth]{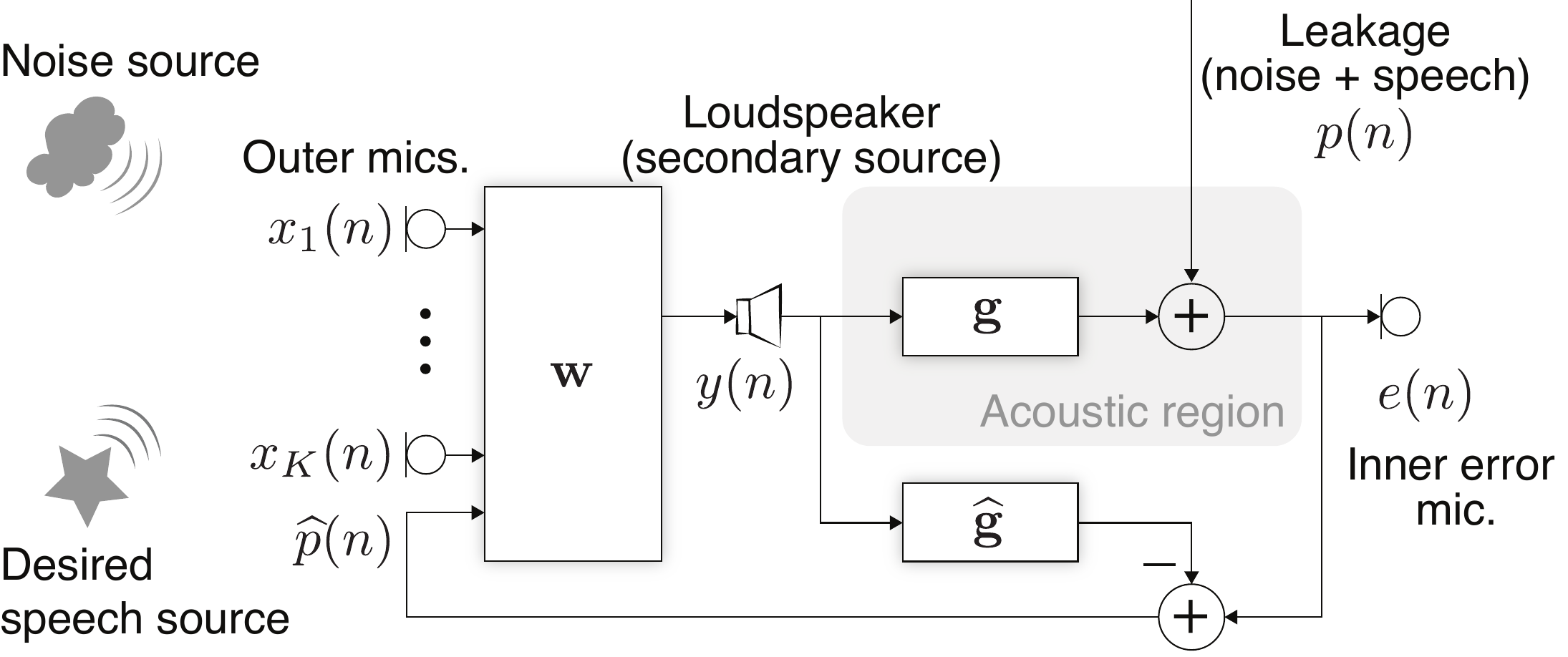}
    \caption{Block diagram of an ANC system with $K$ outer microphones, one inner error microphone and one loudspeaker (i.e., secondary source). The control filter is denoted by $\mathbf{w}$, the secondary path is denoted by $\mathbf{g}$, and its estimate by $\widehat{\mathbf{g}}$.}
    \label{fig:ssanc}
\end{figure}

Active noise control (ANC) hearables create a quiet environment by using secondary sources to generate anti-noise, which minimizes sound at specific positions when superimposed on the primary noise (leakage)~\cite{Elliott2000, Hansen2012active}. Based on their fit, hearables can be categorized as closed-fitting (completely occluding the ear), open-fitting (partially occluding the ear), and open-ear (no occlusion, e.g., smart glasses). Open-fitting and open-ear designs, in particular, reduce the occlusion effect and improve the physical comfort. Recent research focuses on designing intelligent ANC hearables with spatial selectivity, particularly for complex acoustic environments like cocktail-party scenarios involving multiple sound sources from different directions~\cite{kajikawa2012recent, Chang2016Listening, gupta2022augmented}. In these scenarios, listeners may wish to focus on desired sounds from a specific direction (e.g., the front) while blocking out undesired sounds from other directions.

Modern ANC hearables are commonly equipped with multiple microphones, i.e., microphones on the exterior of the hearable and error microphones in the interior close to the eardrum. On the one hand, conventional ANC algorithms suppress all leakage measured by the inner error microphones, including desired speech. On the other hand, multi-microphone noise reduction algorithms perform spatial filtering by reducing all noise while preserving speech from the desired direction~\cite{vanveen1988, Doclo2015, gannot2017consolidated}. However, these algorithms typically ignore the leakage and do not exploit the inner error microphones. Recent advancements have proposed integrating spatial filtering into an ANC system so that sounds from a desired direction are preserved and sounds from undesired directions are actively suppressed~\cite{Serizel2010integrated, Serizel2011output, Dalga2011combined, Dalga2012theoretical, Ho2018integrated, Patel2020design, xiao2023spatial, Zhang2023time, Xiao2024icassp}. A recent study demonstrated that, when the speech component from an outer reference microphone signal is desired, a small delay is required to satisfy causality. The optimal delay is the acoustic delay between the reference microphone and the inner error microphone for the desired source~\cite{Xiao2024icassp}. These findings were derived using a system with \textit{causal} relative impulse responses (ReIRs). However, under more flexible design constraints, \textit{acausal} ReIRs may give a better performance.

In this paper, we present an improved approach to spatially selective ANC (SSANC) that leverages \textit{acausal} ReIRs in the optimization process, thereby substantially improving performance over the causal optimization approach. We analyze the signal distortion, noise reduction, and signal-to-noise ratio (SNR) improvement of the inner error microphone signal when the speech component of an outer microphone signal is desired. We identify both the optimal range of delays and the optimal degree of acausality based on these evaluation metrics. These insights guide the design of a more effective SSANC for open-fitting hearables.

\section{Signal Model}
As shown in~\cref{fig:ssanc}, we consider a hearable with $K$ outer microphones. Without loss of generality, we consider one loudspeaker as the secondary source and one inner error microphone, resulting in a total of $K+1$ microphones. 
We assume that the acoustic feedback paths between the loudspeaker and the outer microphones are known, such that acoustic feedback can be canceled. In addition, we assume that an estimate of the secondary path between the loudspeaker and the inner error microphone is available. Subscripts $(\cdot)_s$ and $(\cdot)_v$ denote the speech and noise components in signals, respectively.

The inner error microphone signal $e(n)$, with $n$ the discrete time index, is given by
\begin{align}
        e(n) &= p(n) + ( \mathbf{G}\mathbf{w})^\mathcal{T}{\mathbf{x}}(n) , \label{eq:e_dGwx}
\end{align}
where $(\cdot )^\mathcal{T}$ denotes the transpose operator. The leakage (including noise and desired speech) at the inner error microphone is denoted by $p(n)$, and the anti-noise component at the inner error microphone is given by $( \mathbf{G}\mathbf{w})^\mathcal{T}{\mathbf{x}}(n)$, where ${\mathbf{w}}$ is the stacked control filter, ${\mathbf{x}}(n)$ is the stacked input vector, and $\mathbf{G}$ represents the secondary path convolution matrix.
The stacked control filter ${\mathbf{w}}$ is defined as
\begin{subequations}
\begin{align}
    \mathbf{w} &= [ \mathbf{w}^\mathcal{T}_1 \;\;\; \mathbf{w}^\mathcal{T}_2 \;\;\; \dots \;\;\; \mathbf{w}^\mathcal{T}_{K+1} \, ]^\mathcal{T} \in \mathbb{R}^{(K+1)L_w} ,
    \\
    \mathbf{w}_k &= \left[{w_{k,0}} \ {w_{k,1}} \ \dots \ {w_{k,{L_w}-1} } \right]^\mathcal{T} \in \mathbb{R}^{L_w},  
\end{align}
\end{subequations}
where $L_w$ denotes the control filter length for each channel. 
The convolution matrix of the secondary path is defined as
\begin{subequations}
\begin{align}
    \mathbf{G} &= \mathrm{blkdiag}\left({\mathbb{G}}  \dots  {\mathbb{G}} \right) \in \mathbb{R}^{(K+1)L \times (K+1)L_w},  \label{eq:G_tilde_multi}
\\ 
    {\mathbb{G}} &= 
        \begin{bNiceMatrix}
                g_0       &  \cdots  & 0         \\[-2pt]
                \vdots    &  \ddots  & \vdots    \\[-2pt]
                g_{L_g-1} &  \ddots  & g_0       \\[-2pt]
                \vdots    &  \ddots  & \vdots    \\[-0pt]
                0         &  \cdots  & g_{L_g-1}
        \end{bNiceMatrix}
        \in \mathbb{R}^{L \times L_w} ,
    \label{eq:G_hat}
\end{align}
\end{subequations}
where $L=L_g+L_w-1$, and $L_g$ denotes the secondary path filter length.
As input signals to the control filter we consider the $K$ outer microphone signals $\mathbf{x}_k(n)$, $k=1, \dots , K$, and an estimate of the leakage $\widehat{\mathbf{p}}(n)$, i.e., the stacked input vector ${\mathbf{x}}(n)$ is defined as
\begin{align}
    {\mathbf{x}}(n) &= [\mathbf{x}_{1}^\mathcal{T}(n) \, \dots \, \mathbf{x}_{K}^\mathcal{T}(n) \,\, \widehat{\mathbf{p}}^\mathcal{T}(n) ]^\mathcal{T} \in \mathbb{R}^{(K+1)L} , \label{eq:x_tilde_multi}
\end{align}
with
\begin{subequations}
\begin{align}
    \mathbf{x}_k(n) &= \left[ x_k(n) \ \dots \ x_k(n-L+1) \right]^\mathcal{T}, \label{eq:x_k_vec}
\\ 
    \widehat{\mathbf{p}}(n) &= \left[ \,\, \widehat{p}\,(n) \, \ \dots \ \,\, \widehat{p}\,(n-L+1) \right]^\mathcal{T}.
\end{align}
\end{subequations}
The estimated leakage $\widehat{p}(n)$ can be computed from the inner error microphone signal $e(n)$, the loudspeaker signal vector $\mathbf{y}(n)=[y(n) \dots y(n-L_g+1)]^\mathcal{T}$, and an estimate of the secondary path $\widehat{\mathbf{g}}$ as
\begin{equation}
\widehat{p}(n)=e(n)-\widehat{\mathbf{g}}^\mathcal{T} \mathbf{y}(n) .
\label{eq:d_hat}
\end{equation}
Assuming a perfect estimate of the secondary path to be available, i.e., $\widehat{\mathbf{g}} = \mathbf{g} = [ {g}_0\ {g}_1\ \allowbreak \dots\ \allowbreak {g}_{L_g-1} ]^\mathcal{T}$, the leakage can be written as $p(n) = \widehat{p}(n) = \mathbf{q}^\mathcal{T}\mathbf{x}(n)$, with 
\begin{subequations}
\begin{align}
    \mathbf{q} &= [ \, \mathbf{0}^\mathcal{T} \ \ldots \ \mathbf{0}^\mathcal{T} \ \bm{\updelta}^\mathcal{T} ]^\mathcal{T} \in \mathbb{R}^{(K+1)L} ,    \label{eq:delta_tilde_multi}
\\
    \bm{\updelta} &= \left[ \, 1 \;\;\;\; 0 \;\;\; \dots \;\;\; 0 \; \right]^\mathcal{T}  \in \mathbb{R}^{L} .
\end{align}
\end{subequations}

Hence, the inner error microphone signal in~\labelcref{eq:e_dGwx} can be rewritten as
\begin{align}
    e(n) = (\mathbf{q} + \mathbf{G}\mathbf{w})^\mathcal{T}{\mathbf{x}}(n) .
    \label{eq:e_qGwx} 
\end{align}

Assuming that the speech and the noise components are uncorrelated, then we have
\begin{subequations}
    \begin{align}
        e_s(n) &= (\mathbf{q} + \mathbf{G}\mathbf{w} )^\mathcal{T} {\mathbf{x}_s}(n), \label{eq:e_s_dGwx}
        \\
        e_v(n) &= (\mathbf{q} + \mathbf{G}\mathbf{w} )^\mathcal{T} {\mathbf{x}_v}(n). \label{eq:e_v_dGwx}
    \end{align}
\end{subequations}

\section{Acausal Optimization}
In the SSANC system, the objective is to minimize the power of the inner error microphone signal while preserving the delayed desired speech component of an outer reference microphone signal. This can be achieved by imposing the constraint
\begin{equation}
    e_s(n) = x_{\mathrm{ref},s}(n-\Delta) .
    \label{eq:es_alpha_x_ref_time}
\end{equation}

For each channel, the speech component can be represented as the convolution of the speech component of the reference microphone signal and an acausal ReIR filter, that is,
\begin{equation}
    x_{k,s}(n) = \sum_{l=-L_a}^{L_h-1} h_k(l) \, {x}_{\mathrm{ref},s}(n-l),  
\end{equation}
where the ReIR has $L_a$-tap anti-causal part and $L_h$-tap causal part. The stacked input vector for all the channels can then be written as 
\begin{equation}
    \mathbf{x}_{s}(n) 
    =  \mathbf{H}^\mathcal{T} \, \mathbf{x}_{\mathrm{ref},s}(n)  ,
    \label{eq:x_s_vec}
\end{equation}
where 
\begin{subequations}
    \begin{align}
        & \hspace{2pt} \mathbf{H} = \left[\mathbf{H}_1  \dots  \mathbf{H}_{K+1} \right] \in \mathbb{R}^{(L_a+L_h+L-1) \times (K+1)L}  ,
        \\[0.2em]
        &\mathbf{H}_k \! =  \!\!
                \begin{bNiceMatrix}
                h_{k,-L_a} & \!\! \cdots   \!\!  & 0               \\[-3pt]
                \vdots     & \!\! \ddots   \!\!  & \vdots          \\[-3pt]
                h_{k,L_h-1}& \!\! \ddots   \!\!  & 0               \\[-3pt]
                0          & \!\! \ddots   \!\!  & h_{k,-L_a}      \\[-3pt]
                \vdots     & \!\! \ddots   \!\!  & \vdots          \\[0pt]
                0          & \!\! \cdots   \!\!  & h_{k,L_h-1}
                \end{bNiceMatrix}
                \! \in \! \mathbb{R}^{(L_a+L_h+L-1) \times L} ,   \raisetag{15pt}
        \\
        &\mathbf{x}_{\mathrm{ref},s}(n) \! = \! [ x_{\mathrm{ref},s}(n+L_a)  \dots  x_{\mathrm{ref},s}(n\!-\!L_h\!-\!L\!+\!2)  ]^\mathcal{T}  \notag
        \\
        & \hspace{35pt} \in \mathbb{R}^{L_a+L_h+L-1} ,
    \end{align}
\end{subequations}
and $\mathbf{x}_{s}(n) \in \mathbb{R}^{(K+1)L}$ is similarly defined as~\labelcref{eq:x_tilde_multi}.

We rewrite~\labelcref{eq:es_alpha_x_ref_time} as
\begin{equation}
    e_s(n) = \bm{\updelta}_{\Delta}^\mathcal{T} \mathbf{x}_{\mathrm{ref},s}(n) ,
    \label{eq:es_alpha_x_ref_tim_rewrite}
\end{equation}
where
\begin{equation}
    \bm{\updelta}_{\Delta} = [ \underbrace{0 \, \dots \, 0}_{L_a} \, \underbrace{0 \, \dots \, 0}_{\Delta} \, \underbrace{1 \,\, 0 \, \dots \, 0  }_{L_h+L-1-\Delta} ]^\mathcal{T} \in \mathbb{R}^{L_a+L_h+L-1}  .  \raisetag{10pt}
    \label{eq:delta_Delta}
\end{equation}
Thus, using~\labelcref{eq:e_s_dGwx,eq:x_s_vec,eq:es_alpha_x_ref_tim_rewrite}, we reformulate \labelcref{eq:es_alpha_x_ref_time} as
\begin{equation}
    \mathbf{H} (\mathbf{q} + \mathbf{G}\mathbf{w} )  = \bm{\updelta}_{\Delta} .
\end{equation}

It is important to note that although the ReIRs, e.g., $\mathbf{H}$ and $\bm{\updelta}_{\Delta}$, are modeled to be \textit{acausal}, the SSANC control filter $\mathbf{w}$ for generating the anti-noise is still \textit{causal}.

Therefore, the optimization problem for an SSANC system can be defined as
\begin{mini!}|l|
    {\mathbf{w}} {\mathcal{E} \left\{e^{2}(n)\right\} + \beta \mathbf{w}^\mathcal{T} \mathbf{w}}
    {}{}
    \addConstraint{\mathbf{H} (\mathbf{q} + \mathbf{G}\mathbf{w} ) }{ = \bm{\updelta}_{\Delta}, }
\end{mini!}
where $\mathcal{E} \{\cdot\}$ denotes the mathematical expectation operator. $\beta$ denotes a regularization factor.
The solution is found to be
\begin{align}
        & \mathbf{w} =    \notag
        \\[0.0em]
        &- \! \left[ \mathbf{I} \! - \! \bm{\Phi}_{\mathbf{rr}}^{-1} \mathbf{G}^\mathcal{T} \! \mathbf{H}^\mathcal{T} \!
        \left( \mathbf{H} \mathbf{G} \bm{\Phi}_{\mathbf{rr}}^{-1} \mathbf{G}^\mathcal{T} \! \mathbf{H}^\mathcal{T} \!\! + \! \rho \mathbf{I} \right)^{-1}  \!
        \mathbf{H} \mathbf{G} \right] \!
        \bm{\Phi}_{\mathbf{rr}}^{-1} \! \bm{\upphi}  \notag
        \\
        &+ \bm{\Phi}_{\mathbf{rr}}^{-1}  \mathbf{G}^\mathcal{T} \! \mathbf{H}^\mathcal{T} \!\!
        \left( \mathbf{H} \mathbf{G} \bm{\Phi}_{\mathbf{rr}}^{-1} \mathbf{G}^\mathcal{T} \! \mathbf{H}^\mathcal{T} \!\! + \! \rho \mathbf{I} \right)^{-1} \!\!
        \left( \bm{\updelta}_{\Delta} \!\! - \mathbf{H} \mathbf{q} \right)  ,
    \label{eq:w_solution}
\end{align}
where
\begin{subequations}
    \begin{align}
        \bm{\Phi}_{\mathbf{xx}} &= \mathcal{E} \left\{{\mathbf{x}}(n) {\mathbf{x}}^\mathcal{T}(n)\right\} ,
        \\
        \bm{\Phi}_{{\mathbf{r}} {\mathbf{r}}} &= \mathbf{G}^\mathcal{T} \bm{\Phi}_{{\mathbf{x}} {\mathbf{x}}} \mathbf{G} + \beta \mathbf{I} ,
        \\
        \bm{\upphi} &= \mathbf{G}^\mathcal{T} \bm{\Phi}_{{\mathbf{x}} {\mathbf{x}}} \mathbf{q} ,
    \end{align}
\end{subequations}
and $\rho$ denotes a small regularization factor, $\mathbf{I}$ denotes the identity matrix.

It should be noted that the solution is similar to the previous studies~\cite{xiao2023spatial, Xiao2024icassp}. The major difference is that the ReIR matrix $\mathbf{H}$ is acausal now containing $L_a$-tap anti-causal part of the ReIR for all channels. When $L_a = 0$, the solution becomes causal, as in the previous approach.

\section{Simulations}
\begin{figure*}[t]
    \centering
    \captionsetup[subfloat]{captionskip=0pt,farskip=0pt}
    \subfloat[]{\includegraphics[height=0.23\textwidth]{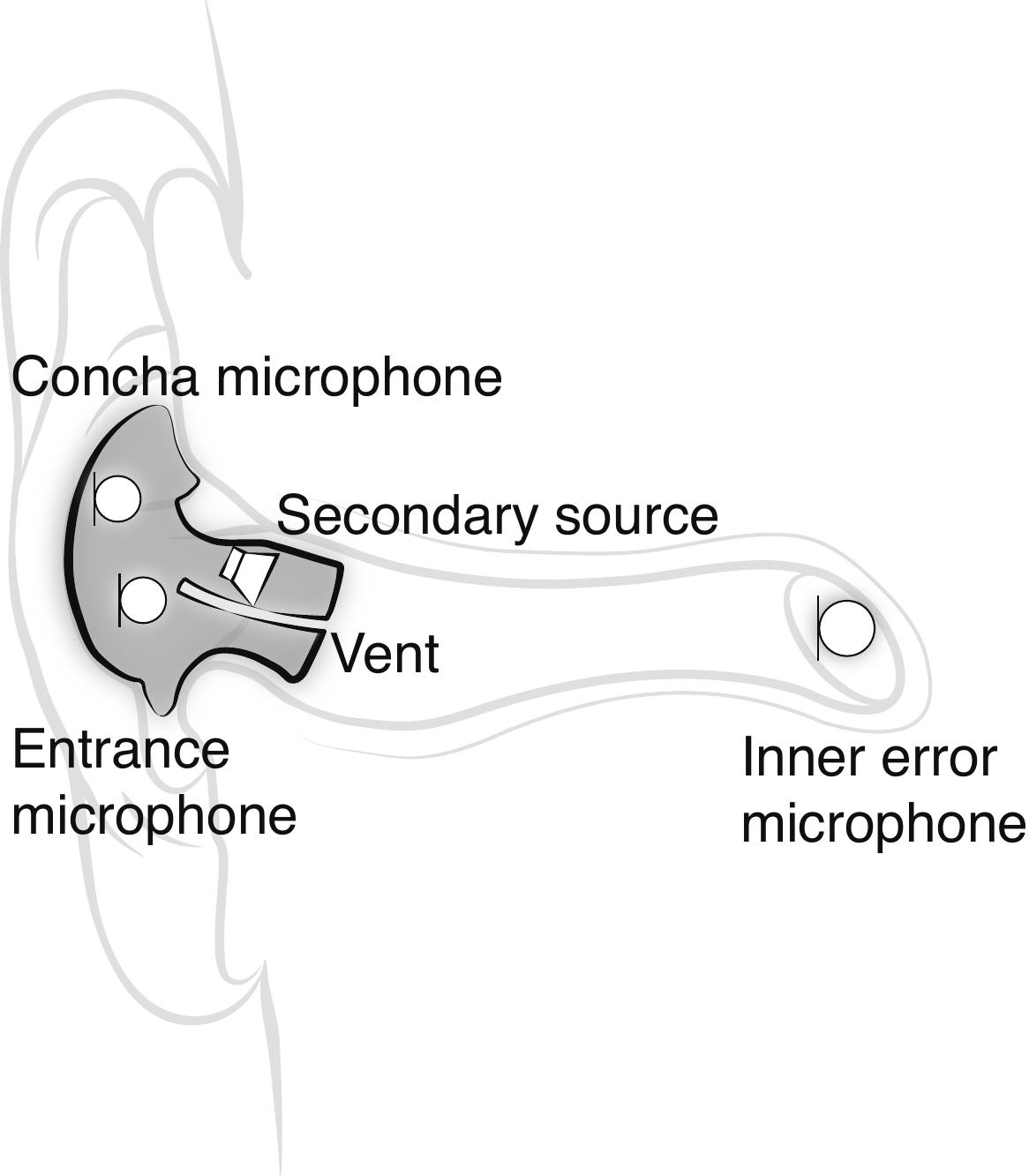} \label{fig:Ear_fa2025}}
    \qquad \qquad 
    \subfloat[]{\includegraphics[height=0.23\textwidth]{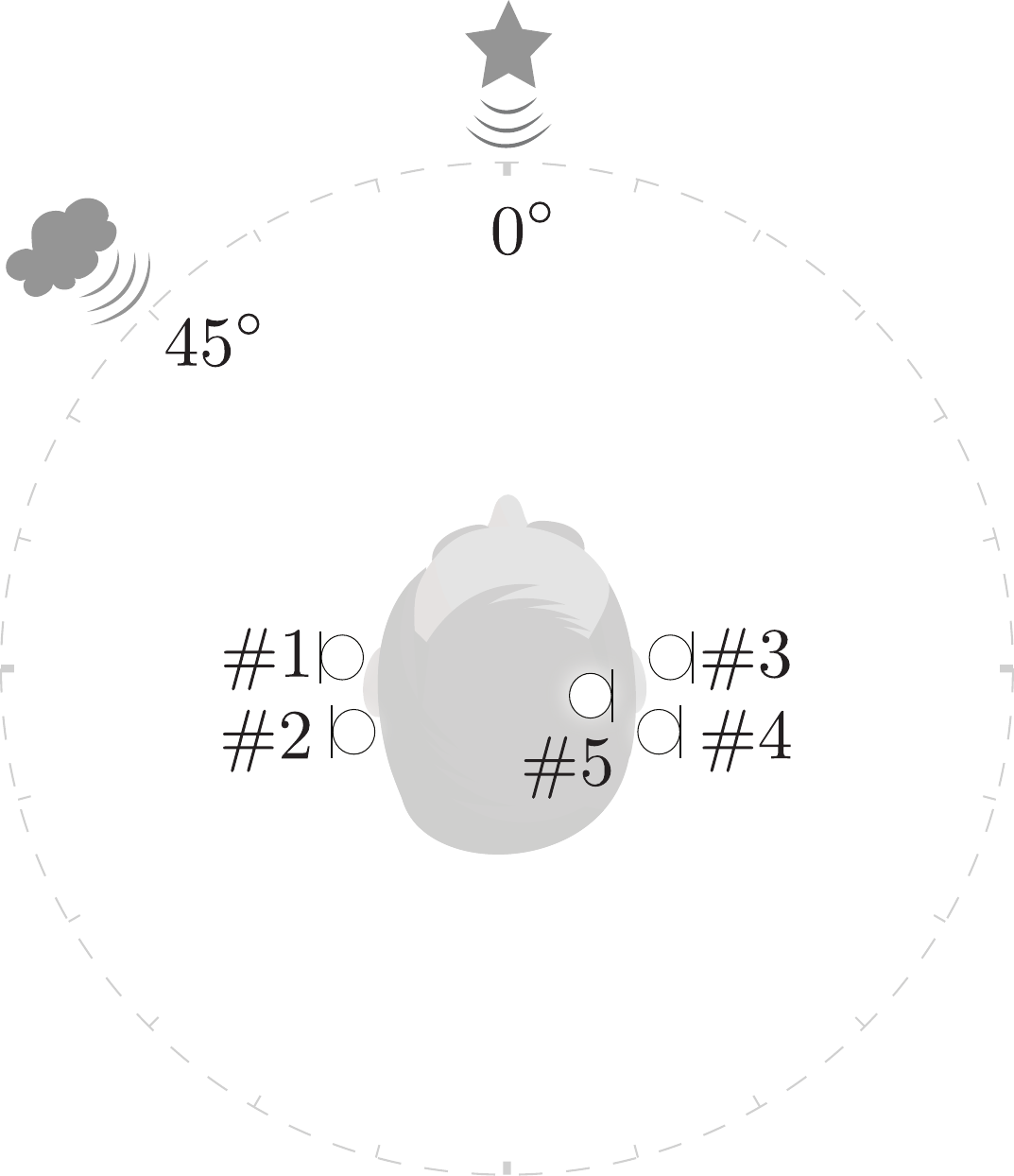} \label{fig:Scenario_1}}
    \qquad \qquad
    \subfloat[]{\includegraphics[height=0.23\textwidth]{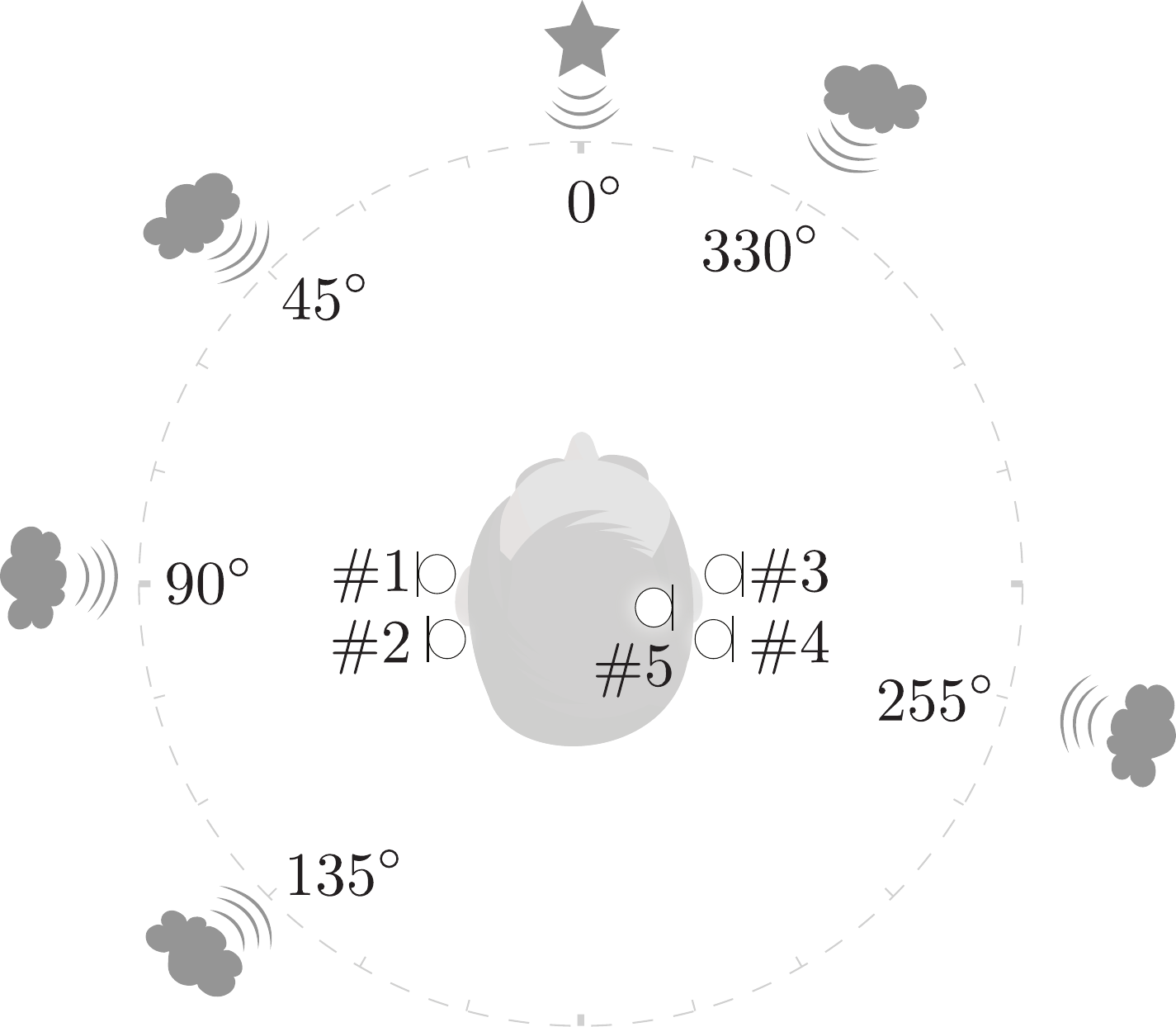} \label{fig:Scenario_2}}
    \vspace{-4pt}
    \caption{(a) Illustration of the open-fitting hearable. (b) First acoustic scenario. A desired speech source is at $0^\circ$, and one undesired speech source is at $45^\circ$. (c) Second acoustic scenario. The desired speech source is at $0^\circ$, and five babble noise sources are at $45^\circ$, $90^\circ$, $135^\circ$, $255^\circ$ and $330^\circ$. }
    \label{fig:Setup}
\end{figure*}
\subsection{Setup}
For the simulation, we considered a pair of open-fitting hearables~\cite{denk2019one,denk2021hearpiece} inserted into both ears of a GRAS 45BB-12 KEMAR Head \& Torso simulator, as shown in~\cref{fig:Ear_fa2025}. We used four outer microphones (entrance microphones and concha microphones at the left and right ears, labeled as \#1-- \#4), one inner error microphone (located at the right ear, labeled as \#5), and the outer receiver at the right ear as the secondary source. The inner error microphone was assumed to be at the eardrum. 

Two acoustic scenarios were considered. In the first scenario, as shown in~\cref{fig:Scenario_1}, we considered a desired speech source at $0^\circ$ and an undesired speech source at $45^\circ$ (\say{p361\_005} and \say{p243\_018} from the VCTK dataset\cite{veaux2017cstr}, respectively). In the second scenario, as shown in~\cref{fig:Scenario_2}, the same desired speech source remained at $0^\circ$, and five noise sources were placed at $45^\circ$, $90^\circ$, $135^\circ$, $255^\circ$ and $330^\circ$ (babble noise from the NOISEX-92 database~\cite{Varga1993assessment}). The desired speech and noise components in all microphone signals were generated by convolving source signals with measured anechoic impulse responses from the database~\cite{denk2021hearpiece}. All signals were 5 seconds in duration and sampled at 16~kHz. The desired speech and noise components were mixed such that the signal-to-noise ratio (SNR) of the leakage at the inner error microphone was set to $-5$~dB. In the second scenario, all noise sources contributed the same energy. As the secondary path estimate we used the measured impulse response between the outer receiver and the inner error microphone at the right ear (from~\cite{denk2021hearpiece}).

We set the filter lengths to $L_w=600$, $L_g=280$, and $L_h=262$. The acausal ReIRs were computed using the least-mean-squares adaptive filter (after convergence) from microphone signals simulated with white noise at $0^\circ$, using the entrance microphone \#3 as the reference microphone. Similar to the previous studies~\cite{xiao2023spatial,Xiao2024icassp}, we were only interested in the frequency range above 100~Hz. Therefore, a minimum-phase high-pass filter with a cut-off frequency at 100~Hz was applied to the desired signal $x_{\mathrm{ref},s}(n-\Delta)$. The effects of delay ($\Delta$) and acausality ($L_a$) on system performance were investigated.

\subsection{Evaluation metrics}
The following three metrics are used for evaluation. The intelligibility-weighted spectral distortion is used to assess the amount of speech distortion~\cite{Spriet2004spatially, Doclo2007frequency, Serizel2010integrated}. It is defined as
    \begin{align}
        &\mathrm{SD}_\mathrm{intellig} \, \mathrm{(dB)}
            = 
            \sum\limits_{b=1}^{B} I(\omega_b) \, 10\log_{10} \frac{\mathcal{P}_\epsilon (\omega_b)}{\mathcal{P}_{\mathrm{ref},s} (\omega_b)}  ,
        \label{eq:SD_intellig}
    \end{align}
where the band importance function $I(\omega_b)$ expresses the importance of the $b$-th one-third octave band for intelligibility~\cite{ASA1997}, and $B$ denotes the total number of bands. $\mathcal{P}_\epsilon (\omega_b)$ is the power spectral density of $\epsilon(n)$ in the $b$-th band, where $\epsilon(n) = e_s(n) - x_{\mathrm{ref},s}(n-\Delta)$. $\mathcal{P}_{\mathrm{ref},s} (\omega_b)$ is the power spectral density of $x_{\mathrm{ref},s}(n-\Delta)$ in the $b$-th band.

The noise reduction is defined as the difference between the power of the noise component of the leakage $p_v(n)$ (without control) and the noise component of the inner error microphone signal $e_v(n)$ (with control), i.e.,
\begin{equation}
    \mathrm{NR} \ \mathrm{(dB)}  
    = 
        10\log_{10}   \sum\limits_{n=1}^{N}  p_v^2(n) 
        -
        10\log_{10}   \sum\limits_{n=1}^{N} e_v^2(n) ,
\end{equation}
where $N$ denotes the total signal length.

The SNR improvement ($\dotDelta \mathrm{SNR}$) is defined as
\begin{align}
  \dotDelta \mathrm{SNR} \ \mathrm{(dB)} 
  \! &= \! 
     10\log_{10} \! \frac{  \sum\limits_{n=1}^{N}  e_s^2(n) }{ \sum\limits_{n=1}^{N} e_v^2(n) }
     -
     10\log_{10} \! \frac{  \sum\limits_{n=1}^{N}  p_s^2(n) }{ \sum\limits_{n=1}^{N} p_v^2(n) }   .
\end{align}

\subsection{Performance for different delays}
\begin{figure*}[t]
    \centering
    \captionsetup[subfloat]{captionskip=0pt,farskip=0pt}
    \subfloat[]{\includegraphics[width=0.47\textwidth]{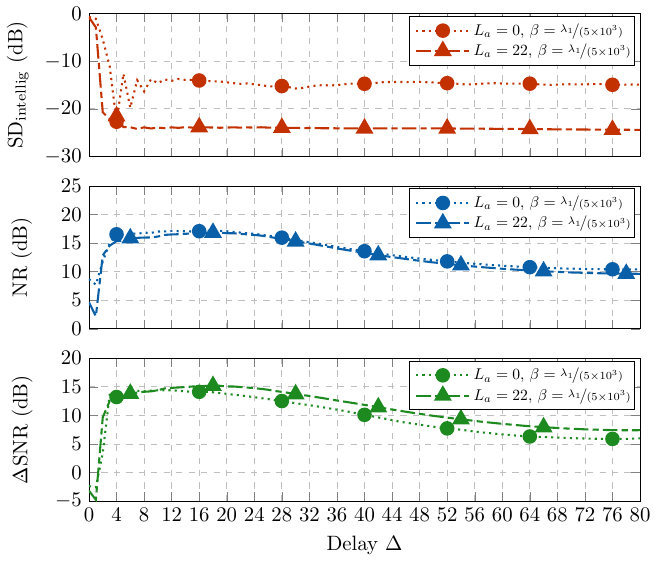} \label{fig:Result_1}}
    \quad  
    \subfloat[]{\includegraphics[width=0.47\textwidth]{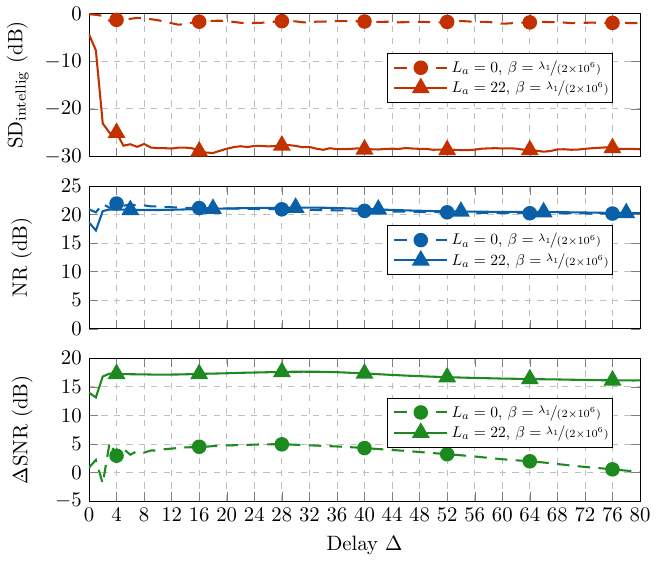} \label{fig:Result_2}}
    \vspace{-10pt}
    \caption{Speech distortion, noise reduction, and SNR improvement for the first simulation scenario for different delay $\Delta$ when (a) $L_a = 0$ and $L_a=22$, with $\beta = \nicefrac{ \lambda_\mathrm{max} (\mathbf{G}^\mathcal{T} \bm{\Phi}_{{\mathbf{x}} {\mathbf{x}}} \mathbf{G} ) }{ (\num{5e3}) } = \nicefrac{ \lambda_1 }{ (\num{5e3}) }$, (b) $L_a = 0$ and $L_a=22$, with $\beta = \nicefrac{ \lambda_1 }{ (\num{2e6}) }$. In all cases, $\rho = \nicefrac{ \lambda_\mathrm{max} (\mathbf{H} \mathbf{G} \bm{\Phi}_{\mathbf{rr}}^{-1} \mathbf{G}^\mathcal{T} \mathbf{H}^\mathcal{T}) }{ (\num{1e5}) }$.   }
    \label{fig:Result}
\end{figure*}

First, we investigate the effect of delay $\Delta$ on performance. The speech distortion, the noise reduction and the SNR improvement are shown in~\cref{fig:Result} for different delays ranging from zero to 80 samples (5~ms) for the first scenario. 

Four design choices are considered.
When $L_a = 0$, $\beta = \nicefrac{\lambda_\mathrm{max} (\mathbf{G}^\mathcal{T} \bm{\Phi}_{{\mathbf{x}} {\mathbf{x}}} \mathbf{G} ) }{ (\num{5e3}) } = \nicefrac{ \lambda_1 }{ (\num{5e3}) }$ and $\rho = \nicefrac{ \lambda_\mathrm{max} (\mathbf{H} \mathbf{G} \bm{\Phi}_{\mathbf{rr}}^{-1} \mathbf{G}^\mathcal{T} \mathbf{H}^\mathcal{T}) }{ (\num{1e5}) }$ as shown in~\cref{fig:Result_1}, where $\lambda_\mathrm{max}(\cdot )$ denotes the largest eigenvalue, this design corresponds to the previous causal optimization~\cite{xiao2023spatial,Xiao2024icassp}. It can be observed that there is very high speech distortion, low noise reduction and low SNR improvement for $\Delta < 4$. This is due to the delays between the reference microphone and the inner error microphone from the desired source, which has a 4-sample delay. The causality constraint requires the delay to be at least four samples. For $4 \leq \Delta \leq 30$, the speech distortion is about $-15$~dB, the noise reduction has about 15--17~dB, and the SNR improvement is about 14~dB. However, further increases in delay result in degraded noise reduction, which in turn leads to reduced SNR improvement. These results agree with the previous findings.

When $L_a = 22$ with the same $\beta$ and $\rho$ as shown in~\cref{fig:Result_1}, it can be observed that the speech distortion is reduced to a much lower level for $\Delta \geq 4$, reaching about $-24$~dB. However, the noise reduction and the SNR improvement do not show significant benefits compared to the previous results.

\begin{figure}[!h]
    \centering
    \includegraphics[width=0.99\linewidth]{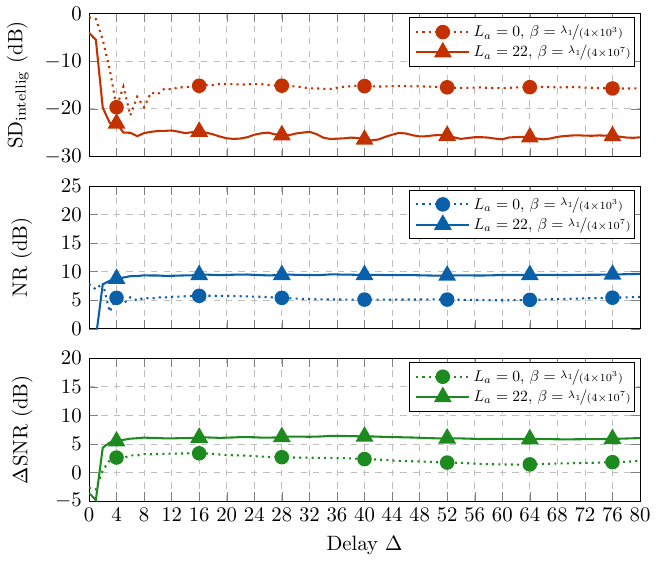}
    \vspace{-10pt}
    \caption{Speech distortion, noise reduction, and SNR improvement for the second simulation scenario for different delay $\Delta$ when $L_a = 0$, $\beta = \nicefrac{ \lambda_1 }{ (\num{4e3}) }$, and when $L_a=22$, $\beta = \nicefrac{ \lambda_1 }{ (\num{4e7}) }$. In both cases, $\rho = \nicefrac{ \lambda_\mathrm{max} (\mathbf{H} \mathbf{G} \bm{\Phi}_{\mathbf{rr}}^{-1} \mathbf{G}^\mathcal{T} \mathbf{H}^\mathcal{T}) }{ (\num{2e5}) }$. }
    \label{fig:Result_5_sources}
\end{figure}

The main advantage of acausal optimization is its ability to achieve better performance with a smaller $\beta$ value, i.e., when less constraint is imposed on the secondary source. The results are shown in~\cref{fig:Result_2} when $\beta = \nicefrac{ \lambda_1 }{ (\num{2e6}) }$. In the causal case ($L_a=0$), the system achieves a high level of noise reduction (21~dB), but at the cost of significant speech distortion (approximately $-2$~dB), resulting in an SNR improvement of just under 5~dB. In this configuration, the SSANC system fails to operate as intended and instead behaves similarly to conventional ANC, reducing both speech and noise components of the leakage. In contrast, the acausal design ($L_a=22$) achieves much lower speech distortion (around $-28$~dB for most delays), maintains a similar level of noise reduction (21~dB), and delivers an SNR improvement exceeding 17~dB for most delays. Furthermore, the acausal approach yields consistent performance across varying delays, reducing the sensitivity to delay selection and thus imposing fewer constraints on the system compared to the causal design, which requires careful tuning of delay parameters.

The results for the second scenario are shown in~\cref{fig:Result_5_sources}. The performance reduces overall compared to the first scenario, which is expected since the scenario is much more challenging. However, the general trend remains the same. The speech distortion is reduced from about $-15$~dB to about $-26$~dB, the noise reduction increases from approximately 6~dB to around $9.5$~dB, and the SNR improvement is increased from about 3~dB to about 6~dB.

\begin{figure*}[t]
    \centering
    \includegraphics[width=0.99\linewidth]{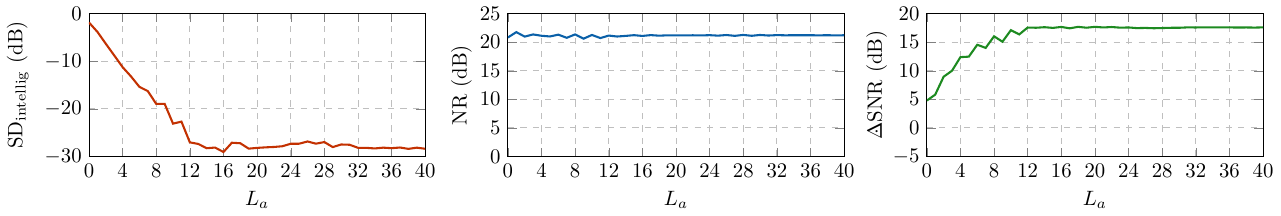}
    \vspace{-8pt}
    \caption{Speech distortion, noise reduction, and SNR improvement for the first simulation scenario for different $L_a$ values when $\Delta = 32$, $\beta = \nicefrac{ \lambda_1 }{ (\num{2e6}) }$ and $\rho = \nicefrac{ \lambda_\mathrm{max} (\mathbf{H} \mathbf{G} \bm{\Phi}_{\mathbf{rr}}^{-1} \mathbf{G}^\mathcal{T} \mathbf{H}^\mathcal{T}) }{ (\num{1e5}) }$.  }
    \label{fig:Result_La}
\end{figure*}
\begin{figure}[t]
    \centering
    \includegraphics[width=0.93\linewidth]{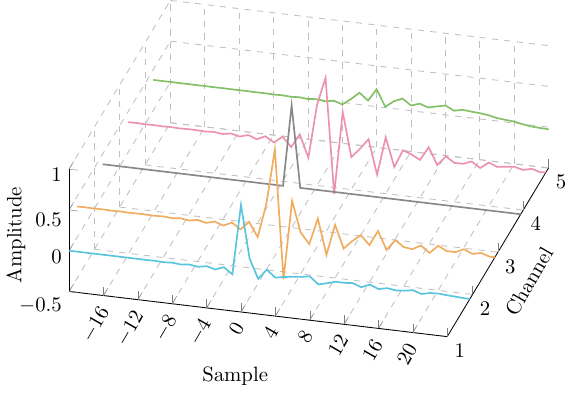}
    \vspace{-9pt}
    \caption{Acausal ReIRs for all five channels.}
    \label{fig:ReIRs}
\end{figure}

\subsection{Performance for different degrees of acausality}
The results above show the performance for the causal design ($L_a = 0$) and one acausal design ($L_a = 22$). We now investigate the effect of different degrees of acausality, i.e., different lengths of anti-causal part $L_a$ on performance. For this analysis, we use the first scenario and fix the delay to $\Delta = 32$ (2~ms), with $\beta = \nicefrac{ \lambda_1 }{ (\num{2e6}) }$ and $\rho = \nicefrac{ \lambda_\mathrm{max} (\mathbf{H} \mathbf{G} \bm{\Phi}_{\mathbf{rr}}^{-1} \mathbf{G}^\mathcal{T} \mathbf{H}^\mathcal{T}) }{ (\num{1e5}) }$. The results for different $L_a$ values are shown in~\cref{fig:Result_La}. 

It can be observed that noise reduction remains consistent across all $L_a$ values. However, speech distortion is high when $L_a$ is small. As $L_a$ increases, the speech distortion decreases and stabilizes when $L_a$ reaches approximately 12. The SNR improvement follows a similar trend.

By examining the ReIRs for all five channels, as depicted in~\cref{fig:ReIRs}, it can be observed that the amplitude of the impulse responses also stabilizes around $L_a = 12$. Thus, using acausal filters to model the response from the desired source enables the SSANC system to achieve better performance than that of a purely causal design.

\section{Conclusion}
This paper has presented an improved approach to spatially selective active noise control that incorporates acausal relative impulse responses into the optimization process, leading to a substantial performance improvement over the causal optimization approach. Using a pair of open-fitting hearables in two acoustic scenarios, we demonstrated that the speech distortion under acausal optimization remains consistently lower than that of the causal optimization. Furthermore, both noise reduction and SNR improvement are significantly higher with acausal optimization. This indicates that acausal relative impulse responses offer a more accurate representation of the desired source, leading to improved noise control performance.

\section{Acknowledgments}
This research was funded by the Deutsche Forschungsgemeinschaft (DFG, German Research Foundation) -- Project-ID 352015383 -- SFB 1330 C1.

\bibliography{main.bib}

\end{document}